\documentclass[prl,twocolumn,floatfix,showpacs]{revtex4}
\usepackage{graphicx}
\usepackage{nccmath}

\begin{document}

\title{Local tensor network for strongly correlated projective states}
\author{B. B\'eri}
\affiliation{Theory of Condensed Matter Group, Cavendish Laboratory, J.~J.~Thomson Ave., Cambridge CB3~0HE, UK}
\author{N. R. Cooper}
\affiliation{Theory of Condensed Matter Group, Cavendish Laboratory, J.~J.~Thomson Ave., Cambridge CB3~0HE, UK}
\date{January 2011}
\begin{abstract}
The success of tensor network approaches in simulating strongly correlated 
 quantum systems crucially depends on whether the many body states that are relevant for the problem can be encoded in a local tensor network. 
Despite numerous efforts, strongly correlated projective states, fractional quantum Hall states in particular, have not yet found a local tensor network representation. Here we show that
 one can encode the calculation of averages of local operators in a Grassmann tensor network which is local. Our construction is explicit, and allows the use of physically motivated trial wavefunctions as starting points in tensor network variational calculations.
\end{abstract}
\maketitle

Approximating ground states of strongly correlated quantum systems is one of the central challenges in condensed matter physics. In recent years there have been several promising proposals aimed at tackling this problem for gapped spin\cite{PEPS,Verstraete2006,Vidal2007,Levin2007a}, as well as fermionic\cite{Corboz2009,Kraus2010,Changlani2009,Corboz2010a,Gu2010} systems with local interactions. Common to these approaches is that they are all based on  a class of states which can be parametrized with  tensor networks.
For a lattice system, the tensors live on the sites, which are connected by a set of links such that the indices of tensors at the ends of the links are contracted (possibly according to a metric defined by link tensors). 
An approximation of the ground state is obtained by treating the tensors as variational degrees of freedom. 
 In order for tensor network approaches to be efficient, it is important that, for an accurate representation of the state, the  number of degrees of freedom per tensor does not grow exponentially with the system size.
Tensor networks with this property are local tensor networks. There exist various schemes for efficiently calculating averages of local operators with respect to local tensor network states\cite{PEPS,Verstraete2006,Vidal2007,Levin2007a,Corboz2009,Kraus2010,Changlani2009,Corboz2010a,Gu2010}.

Strongly correlated systems can realize gapped phases showing topological order, with features (ground state degeneracy, braiding statistics, etc.) that depend only on the topology of the configuration space, not its local details\cite{wen1990topological,WenNiu90}. Due to the topological nature of these phases, it is a nontrivial question whether one can capture topologically ordered ground states with local tensor networks. For spin systems, Refs. \cite{Verstraete2006,Aguado2008,Konig2009,Gu2009a,Buerschaper2009} have shown that the ground states in a class of time-reversal invariant topological phases, so called string-net condensates\cite{Wen2003,Levin2005}, admit a local tensor network representation. The experimentally most accessible topologically ordered states, the fractional quantum Hall (FQH) states, however, are fermionic, break time-reversal symmetry. Their existing tensor network representations in the literature turn out to be nonlocal\cite{Iblisdir2007,Changlani2009,Gu2010}. 
Does this mean that these states, and their spin-system descendants (such as the chiral spin liquid state\cite{KL87,WenWilczekZee}) lie outside of the scope of tensor network based variational simulations?
Fortunately, the results of this Letter demonstrate the contrary: we show that FQH states, or more generally, states from the so-called projective construction\cite{Jain1989,Wen1999} do admit a local tensor network representation. Tensor network based variational schemes should thus be able to find such ground states.

To explain our idea, we start by first describing a non-interacting system: the many electron ground state  $|\Psi\rangle$ of a gapped quadratic Hamiltonian $H=\frac{1}{2}\sum_{kl}(\psi^\dagger_k,\psi_k) h_{kl}(\psi_l, \psi^\dagger_l)^T$, where the indices $k,l$ label sites of a lattice and $\psi^{(\dag)}_l$ are the fermionic field operators on these sites.  (Writing $H$, in terms of a Bogoliubov de Gennes Hamiltonian $h$ with a $2\times 2$ electron-hole structure for each $kl$ is redundant for Hamiltonians without pairing, but it allows us to treat pairing states and states with fixed particle number on equal footing.)  In the usual tensor network approaches, the tensor network representation of $|\Psi\rangle$ would be a preparatory step in expressing  averages  of operators as traces over tensor networks. 
The key observation of our work is that, 
instead of first representing $|\Psi\rangle$, one can construct tensor networks for the averages directly.
 We consider operators of the form $A=\prod_{j\in{K_A}}(\psi_j^\dagger)^{\bar{k}_j}\psi_j^{k_j}$, where   $k_j,\bar{k}_j=0,1$ and $K_A$ is the set of sites $j$ where $A$ acts, i.e.,  where $k_j+ \bar{k}_j\neq 0$. (A general local operator can be obtained as a linear combination of such products.) Using a result of Bravyi\cite{Bravyi2005} we can express $\langle \Psi|A|\Psi\rangle$ as a Grassmann integral, 
\begin{multline}
\langle\Psi|A|\Psi\rangle={\cal N}\int\prod_{j}d\phi_{ j}d\bar{\phi}_{ j}\prod_{jk}e^{-\frac{1}{2}(\bar{\phi}_{j},\phi_{ j})\ h_{jk}^{\text{(fl)}}(\phi_{k},\bar{\phi}_{ k})^{T}} \\ \times \prod_{j\in K_A} \omega[(\psi_j^\dagger)^{\bar{k}_j}\psi_j^{k_j}],       
\label{eq:Bravyi}\end{multline}
where ${\cal N}$ is an $A$-independent normalization factor\cite{Bravyinote}. (We omit ${\cal N}$ henceforth, as it drops from $\langle \Psi|A|\Psi\rangle/\langle \Psi|\Psi\rangle$.) 
Here $\bar{\phi}_j$, $\phi_j$ are Grassmann variables, and $\omega[\psi_j^{(\dagger)}]$ is  $\phi_j/\sqrt{2}$  ($\bar{\phi}_j/\sqrt{2}$) and $\omega[\psi_j^\dagger \psi_j]= \exp(\bar{\phi}_j\phi_j)/2$.
The matrix $h^\text{(fl)}$ is the ``flat band Hamiltonian'', which 
has the same eigenvectors as $h$, but has the eigenvalues $E_j$ replaced by ${\rm sgn}(E_j)$\cite{Kitaev2006,flathnote}. 
Using the Grassmann Leibniz rule (see e.g. Ref~\onlinecite{zinn2002quantum}), Eq.~\eqref{eq:Bravyi} can be transformed from an integral over fermions $\phi_{k}$ living on lattice sites to one over fermions $\phi_{K}$ living on the legs (i.e., ends of links) $K$ of sites $k$,
\begin{equation}
\langle\Psi|A|\Psi\rangle=\ P_{0}\int\prod_{j}T_{j}\prod_{kl}G_{kl}. 
\label{eq:elfnet}
\end{equation}
Here, the symbol $P_0$ represents a projection of the result of the integral to the term containing no Grassmann variables. The factors $G_{kl}=g_{kl}g_{lk}$ with 
\begin{equation}
g_{kl}=\exp[-\frac{1}{2}(\bar{\phi}_{K},\phi_{K})\ h_{kl}^{\text{(fl)}}(\phi_{L},\bar{\phi}_{L})^{T}]
\end{equation}
belong to links $kl$ between sites $k$ and $l$. The precise form of the site factor $T_j$ depends on the operators of $A$ on $j$; in the case that $j\notin K_A$ it is given by
\begin{equation}
T_j=(\sum_{J\in j}d\phi_J)(\sum_{J\in j}d\bar{\phi}_J)-(h_{jj}^\text{(fl)})^{11},
\label{eq:elfT}
\end{equation}
where the indices $11$ refer to the electron-hole structure and $\sum_{J\in j}$ denotes summation over the legs $J$ of site $j$.
Why did we go from Eq.~\eqref{eq:Bravyi} to Eq.~\eqref{eq:elfnet}? Because with Eq.~\eqref{eq:elfnet},  $\langle\Psi|A|\Psi\rangle$ takes the form of a tensor trace in an existing tensor network scheme, namely  the scheme of Grassmann tensor networks introduced by Gu {\it et al.}\cite{Gu2010}. 

Our construction has a saliently pleasant feature: it follows from $h$  explicitly. This is to be compared with the usual situation in tensor network approaches, where if one has an explicit tensor network for a state $|\Psi\rangle$, often it is not derived from a Hamiltonian, but a Hamiltonian (for which $|\Psi\rangle$ is the ground state) is derived from it.
The opposite direction is viable only numerically.
The key feature of our construction, however, is rooted in the fact that for gapped systems, the matrix elements $h^\text{(fl)}_{ij}$ decay exponentially as the function of $|i-j|$\cite{Kitaev2006,Ringel2010}. This implies that our tensor network is \emph{local}. Indeed, introducing a cutoff length $l$, the relative error introduced by neglecting a link longer than $l$ is overestimated by $\exp(-l/\xi)$, where $\xi$ is the decay length of $h^\text{(fl)}$. The total number of links is $N^2$, thus $N^2\exp(-l/\xi)$ largely overestimates the error due the neglected links. Requiring that the error remains fixed implies that we have to scale the cutoff at most as $l\sim \ln N$. The number of legs at site $j$ is estimated as $l^d\sim(\ln N)^d$ in a $d$ dimensional system. Since the number of variables $M$ per leg is independent of the system size, the number of degrees of freedom in $T_j$ (equal to the number of coefficients in its expansion in $d\phi_K$, etc.\cite{Gu2010})  grows at most as $2^{[M(\ln N)^d]}$, i.e., quasi-polynomially.

Solving quadratic Hamiltonians with tensor networks is certainly not a practical strategy. The usefulness of our construction  lies elsewhere: as we now explain, it can be employed to describe an important subset of strongly correlated states, the states from the projective construction. 
These states play an important role in several arenas of strongly correlated quantum systems, such as the FQH effect, spin-liquids, high $T_c$ superconductors, and such novel states of matter as the $d\!=\!3$ fractional topological insulator state\cite{FTI}. 

The projective construction defines strongly correlated trial ground states of the form\cite{Jain1989,Wen1999}
\begin{equation}
\Psi(\{n_j\})=\frac{1}{\sqrt{\prod_{j}n_{j}!}}\langle0|\prod_{j}(\tilde{\psi}_{j})^{n_{j}}|\Theta\rangle.
\label{eq:projgen}
\end{equation}
The state is written in the occupation number representation, with $n_j$ being the occupation of site $j$. The state $|\Theta\rangle$ is the ground state of a gapped quadratic  Hamiltonian $h$ of $P$ flavors of fermions (partons) living in a configuration space identical to the physical one. 
The annihilation operators of the partons are $\{c_{\alpha j}|\alpha=1\ldots P\}$. The state $|\Psi\rangle$ can be bosonic or fermionic. For bosonic (fermionic) states, the operators $\widetilde{\psi}_j$ are even (odd) polynomials of the operators $c_{\alpha j}$. 
The operator $\widetilde{\psi}_j$ should  not be confused with the field operator $\psi_j$: the former acts in the parton Fock space, and is not a field operator, while the latter acts in the physical Fock space, and is a fermionic or bosonic field. 

A well known projective  state is the Laughlin state at filling fraction $\nu=1/3$\cite{LaughlinFQH}. Its wavefunction in position representation is $\Psi(\{z_i\})\sim\Psi_1(\{z_i\})^3$, where $\Psi_1(\{z_i\})$ is the wavefunction of the integer quantum Hall (IQH) state at \mbox{$\nu=1$}. In terms of  Eq.~\eqref{eq:projgen}, $\Psi_3$ corresponds to $P=3$, $\widetilde{\psi}_j=c_{1j}c_{2j}c_{3j}$ and $|\Theta\rangle=|\Psi_1\rangle_1\ |\Psi_1\rangle_2\  |\Psi_1\rangle_3$. 
This Laughlin state is a member of a class of states with\cite{Wen1999}
\begin{equation}
 \tilde{\psi_{j}}=\sum_{\alpha_j=1}^P\chi_{\alpha_{1}\ldots\alpha_{S}}c_{\alpha_{1}j}\ldots c_{\alpha_{S}j},
\label{eq:Zketc}
\end{equation}
where $S\leq P$ an integer. Depending on $|\Theta\rangle$, one can arrive at various FQH states,
such as the $Z_k$ parafermion states\cite{MooreRead,ReadRezayi,Barkesh10}, or the  states based on the composite fermion picture\cite{JainCF}.

To use our tensor network construction for projective states, we need to convert averages of the form $\langle \Psi|A|\Psi\rangle$ into averages with respect to the state $|\Theta\rangle$. (As before, we consider $A=\prod_{j\in{K_A}}(\psi_j^\dagger)^{\bar{k}_j}\psi_j^{k_j}$.) In the bosonic case, we have
\begin{equation}
(\psi_{j}^{\dagger})^{\bar{k}_{j}}\psi_{j}^{k_{j}}|n_{j}\rangle=\frac{\sqrt{n_j!(n_{j}+\bar{k}_{j}-k_{j})!}}{(n_{j}-k_{j})!}|n_{j}+\bar{k}_{j}-k_{j}\rangle
\label{eq:bosonmatrix}
\end{equation}
if $n_j\geq k_j$ and $0$ otherwise. This leads to (with $\bar{k_j}=k_j=0$ for $j\notin K_A$)
\begin{equation}\label{eq:bosonavg}
\langle\Psi|A|\Psi\rangle=
\langle\Theta|\medop\prod_{j}\left[\medop\sum_{n_{j}\geq k_{j}}\frac{(\tilde{\psi}_{j}^{\dagger})^{n_{j}+\bar{k}_{j}-k_{j}}}{(n_{j}-k_{j})!}|0\rangle\langle0|(\tilde{\psi}_{j})^{n_{j}}\right]|\Theta\rangle.
\end{equation}
The key feature is that the expectation value of an operator factorizes into a product over different $j$-s.  This is essential for obtaining a local tensor network in the subsequent stages.  

In the fermionic case, the formula similar to Eq.~\eqref{eq:bosonmatrix} contains a sign factor depending on the occupations of sites $j<l$, 
preventing such a factorization. This can be remedied by using objects with anticommuting ingredients. More precisely, suppose one can find an operator $R_j$ which depends only on partons at $j$, and combines with   $\tilde{\psi}_{j}^{\dagger}$ to mimic field operators in the following sense:
$\{R_j,R_{j^\prime}\}=0$, and for $j^{\prime}\neq j$ $\{R_j,\tilde{\psi}_{j^{\prime}}^{\dagger}\}=0$
and $R_j\tilde{\psi}_{j}^{\dagger}|0\rangle=|0\rangle$,
$R_j|0\rangle=0$. One can then show that
\begin{multline}\label{eq:fermionavg}
\langle\Psi|A(\psi^\dagger,\psi)|\Psi\rangle=\langle\Theta|A(\tilde{\psi}^{\dagger},R) \\ \times \prod_{j}\left[\sum_{n_{j}=0,1}(\tilde{\psi}_{j}^{\dagger})^{n_{j}}|0\rangle\langle0|(\tilde{\psi}_{j})^{n_{j}}\right]|\Theta\rangle,
\end{multline}
where the ordering in the product is arbitrary, since the factors are even in parton operators. Since  $A(\tilde{\psi}^{\dagger},R)$ also factorizes, we now have an average over a product as we wanted. The order of factors matters only for $j\in K_A$ with odd $\bar{k}_j+k_j$. The question is whether  we can find such an $R_j$. We present here the solution of this representation problem for the class of states described by Eq.~\eqref{eq:Zketc}. In that case, $R_j=\tilde{\psi_j}$ satisfies all the requirements, if $\chi$ is chosen to satisfy $\sum_{\{\alpha_l\}}\chi_{\alpha_{1}\ldots\alpha_{S}}\chi^*_{\alpha_{1}\ldots\alpha_{S}}=1/S!$.

Expressions~\eqref{eq:bosonavg},\eqref{eq:fermionavg} give a starting point from which our construction can be built up straightforwardly. Just as in Eq.~\eqref{eq:Bravyi},  we have averages with respect to the ground state $|\Theta\rangle$ of a gapped quadratic fermion Hamiltonian. In both the boson and the fermion case, we have a product between $\langle\Theta|$ and $|\Theta\rangle$ which we denote $\prod_j f_j$. The average will then have the same form as in Eq.~\eqref{eq:Bravyi}, but with $\prod_j\omega(f_{j})$ instead of $\prod_{j\in K_A}\omega[(\psi_j^\dagger)^{\bar{k}_j}\psi_j^{k_j}]$. The factor $\omega(f_{j})$ is obtained similarly as before: one takes $f_{j}(c_{\alpha j}^{\dagger},c_{\alpha j})$, expands it as a normal ordered polynomial in the parton operators. If in a term both $c^\dagger_{\alpha j}$ and $c_{\alpha j}$
are present, they are brought next to each other.  Single factors
$c_{\alpha j}$ ($c_{\alpha j}^{\dagger}$) are then substituted
by $\phi_{\alpha j}/\sqrt{2}$ ($\bar{\phi}_{\alpha j}/\sqrt{2}$),
and the product $c_{\alpha j}^{\dagger}c_{\alpha j}$ by $\exp[\bar{\phi}_{\alpha j}\phi_{\alpha j}]/2$. 

The Grassmann Leibniz rule leads again to an expression for $\langle\Psi|A|\Psi\rangle$ in terms of a tensor network of the form \eqref{eq:elfnet}.
The link factors are now given by $G_{kl}\!=\!g_{kl}g_{lk}$, 
\begin{equation}
g_{kl}\!=\!\exp\left[\!-\frac{1}{2}\medop\sum_{\alpha\beta}(\bar{\phi}_{\alpha K},\phi_{\alpha K})[h_{kl}^{\text{(fl)}}]_{\alpha\beta}(\phi_{\beta L},\bar{\phi}_{\beta L})^T\right],
\label{eq:projG}
\end{equation}
where  $h_{kl}^{\text{(fl)}}$ is a $2P\times 2P$ block between partons at sites $i$ and $j$ of the flat band Hamiltonian associated to $|\Theta\rangle$. 
The Grassmann variables $\bar{\phi}_{\alpha K},\phi_{\alpha K}$ live on the leg of $k$ ending the link $lk$, etc. 

To obtain the site factors, one makes the expansion
\begin{widetext} 
\begin{equation}
\omega(f_{j})\exp\left[-\frac{1}{2}\medop\sum_{\alpha\beta}(\bar{\phi}_{\alpha j},\phi_{\alpha j})[h_{jj}^{\text{(fl)}}]_{\alpha\beta}(\phi_{\beta j},\bar{\phi}_{\beta j})^T\right]=\sum_{\{\pi_{\alpha j},\bar{\pi}_{\alpha j}\}}F_{j}^{\{\pi_{\alpha j},\bar{\pi}_{\alpha j}\}}\prod_{\alpha,\text{inc}}\bar{\phi}_{\alpha j}^{\bar{\pi}_{\alpha j}}\phi_{\alpha j}^{\pi_{\alpha j}},
\label{eq:projTpre}
\end{equation}
where we introduced the expansion coefficients $F^{\{\pi_j,\bar{\pi}_j\}}_j$. The product arranges the $\alpha$-s in increasing order. 
A straightforward calculation leads to 
\begin{equation}
T_{j}=\sum_{\{\pi_{\alpha j},\bar{\pi}_{\alpha j}\}}F_{j}^{\{\pi_{\alpha j},\bar{\pi}_{\alpha j}\}}\prod_{\alpha,\text{inc}}(1-2\pi_{\alpha j})^{1-\bar{\pi}_{\alpha j}}(\sum_{J\in j}d\phi_{\alpha J})^{1-\pi_{\alpha j}}(\sum_{J\in j}d\bar{\phi}_{\alpha J})^{1-\bar{\pi}_{\alpha j}}.
\label{eq:projT}
\end{equation}
\end{widetext}
This tensor network for the states of the projective construction has all the discussed advantages: it is formulated in the framework of an existing scheme, it is explicitly constructed, and most importantly, it is local. Eqs.~\eqref{eq:elfnet}, ~\eqref{eq:projG} and \eqref{eq:projT} are the main results of the Letter.

We now use our results to present the link and site factors for some concrete projective  states. We first consider a fermionic FQH state $|\Psi_{CF}\rangle$ on the lattice from the composite fermion series with Hall conductance $\sigma_{12}\!=\!m/(1+2m)$ (with $m$ integer). It is a generalization of the $\nu\!=\!1/3$ state, obtained by replacing one of the $|\Psi_1\rangle$-s by
$|\Psi_m\rangle$, an IQH state with $\sigma_{12}\!=\!m$\cite{JainCF,KolRead}. 
Its tensor network representation has
\begin{equation}
\!\!\!g_{kl}\!=\!\medop\prod_\alpha\! e^{-H^\alpha_{kl}\bar{\phi}_{\alpha K}\phi_{\alpha L}}, \ H^{1,2}\!=(h^{\text{(fl)}}_1)^{11}\!\!,\ H^3=(h^{\text{(fl)}}_m)^{11}\!,
\end{equation}
where $h^{\text{(fl)}}_m$ is the flat band Hamiltonian obtained from the IQH Hamiltonian $h_m$ realizing the state $|\Psi_m\rangle$.
The site factors for $j\notin K_A$ are given by
\begin{equation}
\!\!T_j\!=\!\frac{1}{4}(\medop\prod_\alpha a_{\alpha j}+\medop\sum_\alpha a_{\alpha j}),\  a_{\alpha j}\!\!=\!\!\!\!\medop\sum_{J,K\in j}\!d\phi_{\alpha J}d\bar{\phi}_{\alpha K}-H^\alpha_{jj}.
\end{equation}
If the operator content of $A$ on site $j$ is $\psi_j$, we have
\begin{equation}
T_j=-\frac{1}{2^{3/2}}\medop\prod_\alpha(\medop\sum_{J\in j}d\bar{\phi}_{\alpha J}).
\end{equation}
The other types of site factors can be found similarly. 

In the CF state, $\tilde{\psi}_j$ was a product of $c$-s. As pointed out in Ref.~\cite{Gu2010}, calculating averages with respect to states with such $\tilde{\psi}_j$ is also possible  by variational Monte Carlo techniques, but these techniques fail once $\tilde{\psi}_j$ is a sum of products of $c$-s. It is thus important that our construction includes the latter case as well. As an illustration we consider (the lattice version of) the bosonic Pfaffian state. In terms of Eq.~\eqref{eq:projgen}, this state has $|\Theta\rangle\!=\!|\Psi_1\rangle |\Psi_1\rangle|\Psi_1\rangle |\Psi_1\rangle$ and
$\tilde{\psi}_j\!=\!c_{1j}c_{4j}\!-\!c_{3j}c_{2j}$\cite{Wen1999}.
 Its tensor network representation has
\begin{equation}
g_{kl}=\medop\prod_{\alpha=1}^4 e^{-H^\alpha_{kl}\bar{\phi}_{\alpha K}\phi_{\alpha L}},\quad H^\alpha=(h^\text{(fl)}_1)^{11}.
\end{equation}
The site factors, for example for $j\notin K_A$, are given by
\begin{multline*}
T_j=\frac{1}{16}[\medop\prod_\alpha (a_{\alpha j}+1)+4(a_{1j} a_{4j}+1)(a_{2j} a_{3j}+1)\\
-4\medop\sum_{J,K,L,M\in j}(
d\bar{\phi}_{1 J}d{\phi}_{2 K}d{\phi}_{3 L}d\bar{\phi}_{4 M}
+
d{\phi}_{1 J}d\bar{\phi}_{2 K}d\bar{\phi}_{3 L}d{\phi}_{4 M}
)
].\end{multline*}

Once the site and link factors $T_j$ and $G_{kl}$ are known, the calculation of $\langle\Psi|A|\Psi\rangle$ amounts to the evaluation of the tensor trace Eq.~\eqref{eq:elfnet}. In tensor network approaches, evaluating tensor traces is a challenge by itself, which requires approximations. Depending on the concrete tensor network framework at hand, one can choose from existing approximation schemes to tackle this task\cite{PEPS,Verstraete2006,Vidal2007,Levin2007a,Corboz2009,Kraus2010,Changlani2009,Corboz2010a,Gu2010}. As Eq.~\eqref{eq:elfnet} is formulated entirely in terms of the framework of Ref.~\onlinecite{Gu2010}, it can be subjected to the approximation scheme developed there, the Grassmann tensor entanglement renormalization group. The concrete implementation of this renormalization procedure for our networks is however beyond the scope of this Letter. 

In conclusion, we have demonstrated that strongly correlated projective states  can be represented using local Grassmann tensor networks.  The results are valid for the whole scope of the projective construction, which include FQH states, spin-liquids, fractional  topological insulators, etc. 
On the level of principles, our work shows that strongly correlated quantum systems where the ground state is close to a projective state can be meaningfully approached using tensor network based algorithms.
From a computational point of view, our development brings the prospects of variational simulations with projective trial states qualitatively closer: first, it breaks an exponential bound faced in previous efforts in representing such states via tensor networks; and, second, it does so without giving up on the explicit nature of the representation. This allows to use physically motivated trial wavefunctions as starting points, which could speed up variational calculations dramatically. 

This work was supported by EPSRC Grant EP/F032773/1.

\end{document}